# Electrical Manipulation of a Topological Antiferromagnetic State

†Hanshen Tsai[1,2], †Tomoya Higo[1,2], Kouta Kondou[2,3], Takuya Nomoto[2,4], Akito Sakai[1,2], Ayuko Kobayashi[1], Takafumi Nakano[2,5], Kay Yakushiji[2,5], Ryotaro Arita[2-4], Shinji Miwa[1,2,6], YoshiChika Otani[1-3,6], and Satoru Nakatsuji[1,2,6,7,*]

[1] *Institute for Solid State Physics, University of Tokyo, Kashiwa, Chiba 277-8581, Japan*

[2] *CREST, Japan Science and Technology Agency, Kawaguchi, Saitama 332-0012, Japan*

[3] *RIKEN, Center for Emergent Matter Science (CEMS), Saitama 351-0198, Japan*

[4] *Department of Applied Physics, University of Tokyo, Bunkyo-ku, Tokyo 113-8656, Japan*

[5] *National Institute of Advanced Industrial Science and Technology (AIST), Spintronics Research Center, Tsukuba, Ibaraki 305-8568, Japan*

[6] *Trans-scale Quantum Science Institute, University of Tokyo, Bunkyo-ku, Tokyo 113-0033, Japan*

[7] *Department of Physics, University of Tokyo, Bunkyo-ku, Tokyo 113-0033, Japan*

†These authors contributed equally to this work.

**Electrical manipulation of emergent phenomena due to nontrivial band topology is a key to realize next-generation technology using topological protection. A Weyl semimetal is a three-dimensional gapless system that hosts Weyl fermions as low-energy quasiparticles[1-4]. It exhibits various exotic phenomena such as large anomalous Hall effect (AHE) and chiral anomaly, which have robust properties due to the topologically protected Weyl nodes[1-23]. To manipulate such phenomena, the magnetic version of Weyl semimetals would be useful as a magnetic texture may provide a handle for controlling the locations of Weyl nodes in the Brillouin zone. Moreover, given the prospects of antiferromagnetic (AF) spintronics for realizing high-density devices with ultrafast operation[24-26], it would**

**be ideal if one could electrically manipulate an AF Weyl metal. However, no report has appeared on the electrical manipulation of a Weyl metal. Here we demonstrate the electrical switching of a topological AF state and its detection by AHE at room temperature. In particular, we employ a polycrystalline thin film[27] of the AF Weyl metal $Mn_3Sn$[10,13,15,28], which exhibits zero-field AHE. Using the bilayer device of $Mn_3Sn$ and nonmagnetic metals (NMs), we find that an electrical current density of ~$10^{10}$-$10^{11}$ A/m$^2$ in NMs induces the magnetic switching with a large change in Hall voltage, and besides, the current polarity along a bias field and the sign of the spin Hall angle $\theta_{SH}$ of NMs [Pt ($\theta_{SH} > 0$)[29], Cu($\theta_{SH} \sim 0$), W ($\theta_{SH} < 0$)[30]] determines the sign of the Hall voltage. Notably, the electrical switching in the antiferromagnet is made using the same protocol as the one used for ferromagnetic metals[31,32]. Our observation may well lead to another leap in science and technology for topological magnetism and AF spintronics[33].**

Recent extensive studies in condensed matter physics have led to the discoveries of novel quantum phases with nontrivial topology in the electronic band structure[4,34,35]. One example of such topological systems is a Weyl semimetal[1-4]. Two non-degenerate bands linearly touch at a pair of momentum points, forming gapless Weyl fermions with different chiralities in a time-reversal-symmetry (TRS) or inversion-symmetry breaking state. These touching points or Weyl nodes act as topologically protected, unit-strength (anti)monopoles of underlying Berry curvature, and lead to various emergent phenomena such as large AHE, anomalous Nernst effect (ANE), chiral anomaly and optical gyrotropy[1-23].

For developing science and technology utilizing novel topological states, a crucial next step would be to manipulate these emergent phenomena electrically. In a Weyl semimetal, the manipulation can be made by moving the Weyl points around in the Brillouin zone. For this purpose, the TRS breaking or magnetic Weyl semimetals are suitable as their magnetic texture may provide a handle for the manipulation. Besides, antiferromagnets (AFMs) have recently attracted significant attention as the active material for next-generation spintronics devices, with the prospects for a higher density storage and much faster operation speed than ferromagnetic counterparts[24-26]. However, there have been no reports on electrical manipulation of either AF or ferromagnetic Weyl semimetals.

Meanwhile, advance in the understanding of topological aspects in electronic structure has led to the discovery of AHE[36-38] in non-ferromagnetic systems such as spin liquids[39] and chiral AFMs $Mn_3X$ ($X$ = Sn, Ge, Ga, and Ir, Pt, Rh)[10,40-44]. The discovery has shown that the AFMs may exhibit large transverse responses such as AHE and ANE due to the Berry curvature in the momentum space in the absence of magnetization $M$[10,40-44]. In particular, the theoretical studies as well as experiments based on single crystals have clarified that $Mn_3Sn$ hosts magnetic Weyl fermions[10,13,15,28]. Associated with the topological protection of Weyl nodes, the large topological responses such as AHE and ANE are robust against disorder, impurity, and thermal fluctuation. For example, they appear over a wide range of Mn concentration[10,13,15] and even in a polycrystalline thin film[27,45], paving the way for future applications.

Here, we demonstrate the electrical switching of the topological AF state and the detection by AHE in a polycrystalline thin film of $Mn_3Sn$. Our work is significant not only for the first successful electrical manipulation of AHE in a topological Weyl metal but as it introduces the same switching protocol as the one used for the conventional spintronics[31,32]. $Mn_3Sn$ is one of the best-studied among various kagome-based metals that have attracted significant interest due to their nontrivial band topology[10,19,20,23,46] and novel magnetic responses at interfaces[47-49]. Besides, the effect of spin injection into $Mn_3Sn$ has been theoretically predicted, promoting its application in spintronics[48,49].

Mn$_3$Sn has the hexagonal D0$_{19}$ structure with the ABAB-stacking of a (0001)-kagome layer of Mn, and the geometrical frustration leads to a three-sublattice non-collinear AF ordering of Mn spins below the Néel temperature $T_N$ ~ 430 K[10,50] (Fig. 1a). Particularly, this AF spin texture can be viewed as a ferroic ordering of a cluster magnetic octupole $T$ (Fig. 1b left)[51], and this ferroic octupole order allows the $M$-linear responses to appear, such as AHE[10], ANE[13], and magneto-optical Kerr effect (MOKE)[52], instead of a very tiny uncompensated $M$~0.006 $\mu_B$/f.u. induced by the spin canting within the (0001)-plane. Strikingly large AHE and ANE in Mn$_3$Sn, comparable with or even exceeding those in ferromagnets (FMs), come from the Weyl fermions and the large Berry curvature, which is equivalent to a few 100 T[13,15,28]. Moreover, the polarization direction of the magnetic octupole determines the location of Weyl nodes and the associated distribution of the Berry curvature in the momentum space (Fig. 1b right)[15]. Thus, it is essential to control the orientation of the magnetic octupole for the electrical manipulation of the topological responses.

Before going to the discussion on the electrical switching measurements, let us first demonstrate that our thin films host the same topological Weyl semimetal state as in the Mn$_3$Sn single crystals[13,15,28]. For this, we employ two types of transport probes for the Weyl fermions, (1) chiral anomaly and (2) ANE [1,4,6,7,9,11-18,21-23] (Methods). First, our measurements using thin films find magneto-transport phenomena fully consistent with the chiral anomaly due to the Weyl fermions (Extended Data Figs. 1 and 2). For instance, the angular dependence of the magnetoconductivity $\sigma$ and planar Hall conductivity $\sigma_H^{PHE}$ are well fit by the theoretical equations for the chiral anomaly[16,17,21,22] (Fig. 1c). Second, ANE probes the Berry curvature at the Fermi energy $E_F$[38]. Thus, the Weyl points near $E_F$ could enhance ANE much more than what would be expected based on the empirical scaling law with magnetization known for FMs[13,14,18,23]. Our measurements using the Mn$_3$Sn thin films reveal that this is indeed the case (Extended Data Figs. 3 and 4a), confirming the magnetic Weyl semimetal state realized in the thin films. Besides, we find that the associated transverse thermoelectric conductivity $\alpha_N$ changes its sign by reversing magnetic field, fully overlapping on top of the field dependence of the Hall conductivity (Extended Data Fig. 4b). This further demonstrates that similar to $\alpha_N$, the switching of AHE well detects the rotation of the pairs of the Weyl points and the associated Berry curvature in the momentum space (Extended Data Fig. 5).

To generate a spin current by the spin Hall effect (SHE), we employ nonmagnetic metals (NM = Pt, W, Cu) in the device with the structure Si/SiO$_2$/Ru(2)/Mn$_3$Sn(40)/NM($d_{NM}$)/AlO$_x$(5) (in nm) deposited on a Si/SiO$_2$ substrate (Methods, Figs. 1d and 1e). First, we measure the Hall voltage $V_H$ as a function of the out-of-plane magnetic field $H_z$ to quantitatively estimate the population of switchable domains normal to the plane of the Mn$_3$Sn layer in the device. Figure 1f shows the clear hysteresis of the Hall voltage with the zero-field change of $\Delta V_H^{field}$ (= $V_H(+H_z\rightarrow 0) - V_H(-H_z\rightarrow 0)$)[27]. As the polarization direction of the octupole points to the same direction as the tiny canted moment, the results indicate that the negative (positive) value of $V_H$ is generated by the "+z (−z) domain" with the positive +z (negative −z) component of the polarization direction of the octupole.

Now, let us examine the possible electrical switching for the bilayer device. A 100 ms write-pulse current $I_{write}$ followed by DC read current $I_{read}$ = 0.2 mA is applied along the *x*-direction (Fig. 1d, Extended Data Fig. 6a). Here, the current $I$ flows in the entire stack, and the current density $J$ is for the part flowing in the NM layer. Remarkably, the electrical current that flows in the device results in different signs of the electrical switching according to the sign of the spin Hall angle $\theta_{SH}$ for the NM layer (Fig. 1g). As for NM = Pt ($\theta_{SH} > 0$)[29], a clear negative (positive) jump appeared in the Hall voltage under a positive (negative) current larger than a critical threshold write-current $I_c$ in a bias field $H_x$ along the current direction $x$. The magnitude of the jump reaches ~30 % of the total Hall voltage change $|\Delta V_H^{field}|$ in the field sweep measurements. As W has a large $\theta_{SH}$ with an opposite sign ($\theta_{SH} < 0$) to Pt[30], we carry out the switching experiments by replacing Pt by W (Fig. 1d). Significantly, the W device exhibits the switching polarity opposite to the Pt one, and the magnitude of the voltage jump amounts to ~25 % of $|\Delta V_H^{field}|$ (Fig. 1g). We have also fabricated the Cu device (Fig. 1g) and a device without the NM layer with the same configuration and found almost no hysteresis with the electrical current cycle (Extended Data Figs. 7a-c). The difference in the switching polarity between the Pt and W devices and the absence of the switching in the Cu and no-NM layer devices cannot be accounted for by the Oersted field generated by the electrical current and the spin-orbit torque (SOT) due to SHE in the $Mn_3Sn$ layer, but agree well with the sign of $\theta_{SH}$ of the NM layer. These results demonstrate that SOT from SHE of the NM layer induces the perpendicular switching of the AF domain.

Generally, the sign of the SOT switching is determined by the bias field along the $I_{write}$ direction. To examine this, under various directions of the bias field, we measure the Hall voltage $V_H$ as a function of $I_{write}$ ($V_H$-$I_{write}$ loop). Figure 2a clearly shows that if the directions of both $I_{write}$ and $H_x$ are the same, the voltage exhibits a negative jump at $|I_c|$ accompanied by the increase in the population of the "+z domains" with the positive (+) z-direction component of the cluster magnetic octupole. If the directions are opposite to each other, the jump becomes positive, increasing the population of the "−z domains". In addition, the magnitude of the jump, $\Delta V_H^{current} = V_H(+I_{write}\rightarrow 0) - V_H(-I_{write}\rightarrow 0)$, changes with the bias field along the current ($x$) direction (Figs. 2b and 2c). The same switching measurements performed under bias magnetic fields ∥ $y$ and $z$ indicate that only the bias field ∥ $x$ induces the switching of the AF domains. This observation follows the expectation of the symmetry requirement of the SOT switching of the perpendicular magnetization[31,32]. The NM thickness dependence of the switching Hall voltage $|\Delta V_H^{\text{current}}|$ in the Pt and W devices shows a systematic change over 20 pieces of devices with a saturation ~ 30 % of $|\Delta V_H^{\text{field}}|$ at $d_{\text{NM}} \geqq 2$ nm, indicating the robust switching properties in a wide range of the NM thickness (Fig. 2d, Extended Data Fig. 6b).

Significantly, the switching or critical write current density $J_c$ is found reasonably small. It is estimated to be $2 \times 10^{11}$ A/m$^2$ for 7.2 nm thick Pt and $5 \times 10^{10}$ A/m$^2$ for 7.2 nm thick W devices. These are smaller than the original values reported for the first observations of the electrical switching in the NM/FM devices ($J_c \sim 10^{12}$ A/m$^2$)[31,32], and comparable to recent estimates for AFM/FM devices (AFM layer as a spin current source)[53], the collinear Néel-SOT devices[54-56], and collinear AFM/Pt devices[57-59]. The estimated heating of the central part of the device is ~ 50 K, and the temperature remains lower than the Néel point even when the write current $I_{\text{write}}$ is on. Moreover, $V_H$ is not affected by heating effect due to the $I_{\text{write}}$ injection because of the wait time of 600 ms (Extended data Figs. 6a, 7d, Methods). We have also carried out the same switching measurements at lower temperatures than room temperature such as 200, 250, 295 K (Fig. 2e). Notably, the same switching takes place, and the threshold current increases only slightly. All experiments confirm that our bilayer devices exhibit the deterministic magnetic switching due to SOT exerted on the AF spin texture. This switching can also support reproducible bipolar writing as an AF memory. Alternate pulse currents with a different sign systematically produce the magnetic switching of the reading $V_H$ over 200 times (Fig. 3a) and the signal is stable against the consecutive injections of $I_{\text{write}} > I_c$ (Extended Data Fig. 7e), demonstrating its controllability.

To understand the deterministic switching mechanism based on SOT, we study the dynamics of the sublattice moments $\boldsymbol{m}_{ia}$ on the one-layer kagome lattice (one blue layer on the $yz$-plane in Fig. 4a) obeying the Landau-Lifshitz-Gilbert (LLG) equation[29,60]

$$\dot{\boldsymbol{m}}_{ia} = -|\gamma|\boldsymbol{m}_{ia} \times \boldsymbol{H}_{\text{eff},ia} + \alpha \boldsymbol{m}_{ia} \times \dot{\boldsymbol{m}}_{ia} + \boldsymbol{T}_{ia} \quad (1)$$

where the suffix $i$ denotes a unit cell, $a = (1, 2, 3)$ a sublattice. Here, the effective magnetic field $\boldsymbol{H}_{\text{eff},ia}$ is defined as $\boldsymbol{H}_{\text{eff},ia} = -M_S^{-1}\delta\mathcal{H}/\delta\boldsymbol{m}_{ia}$ with $M_S = 3$ μ$_B$ of a saturation magnetic moment of Mn atom (Methods). The first, and second terms of r.h.s. represent respectively gyroscopic torque, and Gilbert-damping torque. $\gamma$ ($< 0$) represents the gyromagnetic ratio of an electron, $\alpha$ the Gilbert-damping coefficient. The third term represents the external torque due to the spin injection, namely, the spin-transfer-induced in-plane torque, $\boldsymbol{T}_{ia} = \frac{\hbar|\gamma|J_0\theta_{\text{SH}}}{2em_Sd}\boldsymbol{m}_{ia} \times (\boldsymbol{y} \times \boldsymbol{m}_{ia})$ caused by SOT. Here, $\theta_{\text{SH}}$ denotes the spin-Hall angle in Pt[29], $d$ the thickness of Mn$_3$Sn layer, $m_S = 6M_S$ /unit cell volume, $J_0$ the current density, and $\boldsymbol{y} = (0,1,0)$ the unit vector along $y$-axis.

Given the polycrystalline character of our Mn$_3$Sn layer, there are three crystal-grain configurations representing the Mn$_3$Sn Hall bar devices (Methods, Extended Data Figs. 8 and 9). Here, we focus on the configuration (a) the kagome-layer $\perp$ $I$ and $\parallel$ $\boldsymbol{p}$ (Fig. 4a), where $\boldsymbol{p}$ represents the polarization direction (and the direction anti-parallel to that of spin angular momentum) of spin accumulation at the Mn$_3$Sn/Pt interface, as we find that the deterministic switching accompanied by the AHE change may arise only in this configuration (Methods). For the simulation of the magnetization switching process, we set parameters consistent with the experimentally obtained physical parameters in Mn$_3$Sn, where the six-fold magnetocrystalline anisotropy energy and the spontaneous magnetization are 310 J/m$^3$ and 0.01 μ$_B$/f.u., respectively[61]. Figure 4b shows an example of numerical results for the evolution of an in-plane angle $\varphi$ of the octupole polarization $T$ as a function of time $t$ for the case where $\boldsymbol{p} \parallel y$ and $\boldsymbol{H} \parallel x$. Here, $\varphi$ refers to the angle within the kagome plane ($yz$-plane in Fig. 4a) measured from the $y$-direction ($\varphi = 0$) parallel to $\boldsymbol{p}$ (Fig. 4c, Extended Data Fig. 9a). In Fig. 4b, we show the results calculated with the initial spin structure where the octupole polarization points to $\varphi = \pi/6$, one of the easy axis directions.

Let us illustrate the switching dynamics schematically in Figs. 4d-4f. Before turning on the write current $I_{\text{write}}$, we set the initial spin state to have $\varphi = \pm\pi/6$ as shown in Fig. 4d. Subsequently, the application of $I_{\text{write}}$ exerts in-plane spin-transfer torques $\boldsymbol{T}_{ia}$. As a consequence, torque in the form of $\boldsymbol{T}"_{ia} = \alpha \boldsymbol{m}_{ia} \times \boldsymbol{T}_{ia}$, is exerted on the sublattice moments $\boldsymbol{m}_{ia}$ in the direction of $\boldsymbol{m}_{ia} \times \boldsymbol{p}$, and thus have out-of-plane components. Unlike the SOT switching of the ferromagnetic magnetization, where $\boldsymbol{T}_{ia}$ dominates, the rotational motion of the octupole polarization $T$ is determined by the sum of the precessional movements of the sublattice moments $\boldsymbol{m}_{ia}$ driven by $\boldsymbol{T}"_{ia}$. The spin-torque leads to the steady state under $I_{\text{write}}$ by rotating the octupole polarization to $\varphi = 0 \pm \delta$ during the application of the write current $I_{\text{write}}$. In the absence of the bias field, the two initial states converge to the same steady state irrespective of the sign of $I_{\text{write}}$ (Figs. 4d-4e, Extended Data Fig. 10a, Methods).

Upon turning off $I_{\text{write}}$, the octupole-polarization direction is relaxed and finally aligned along the closest easy-axes ($a$-axes) || $\{2\bar{1}\bar{1}0\}$ selected by the sign of $\pm\delta$, and thus by the bias field directions. More precisely, the sign of $\pm\delta$ is determined by the torque exerted on the $c$-axis-components of the sublattice moments $\boldsymbol{m}_\perp$ induced by the bias field $H_x$ ($\boldsymbol{m}_\perp \times \boldsymbol{p} \parallel \boldsymbol{H} \times \boldsymbol{p}$) (Fig. 4e). As a result, the $I_{\text{write}}$ sweep induces the switching of the octupole polarization $T$ between $b$: $\{01\bar{1}0\}$-axes inclined by $(\pm)\pi/6$ from the horizontal direction to the interface (Fig. 4f). This is analogous to the ferromagnetic case; using a perpendicular magnetized ferromagnetic film, SOT may cause the deterministic switching between two closest stable directions concerning the anisotropy energy, i.e., perpendicular directions. In our model, it corresponds to $\varphi = \pm\pi/6$ along the easy-axes || $\{2\bar{1}\bar{1}0\}$ (Fig. 4f), and we have confirmed that this is always the case with realistic numerical conditions. As the steady-state under $I_{\text{write}}$ does not depend on the polarity of $I_{\text{write}}$ and the final state is determined by the torque $\sim \boldsymbol{m}_\perp \times \boldsymbol{p} \parallel \boldsymbol{H}$

× $p$, the same deterministic switching takes place between another pair of the easy axes, $\varphi = \pm 5\pi/6$.

If we focus on the initial and final states alone, the deterministic switching of the octupole polarization $T$ in Mn$_3$Sn is overall analogous to that of the magnetization in FMs because the same symmetry requirements apply to both cases. However, we note that the emergence of the same steady states irrespective of the sign of $I_{\text{write}}$ (Fig. 4e) is unique to our case as the details of the switching dynamics are different from the magnetization switching in FMs, where the steady states depend on the sign of $I_{\text{write}}$ through the anti-damping SOT, $T_{ia}$ (Methods, Extended Data Fig. 10b). Besides, our results indicate that the switching leads to the change in AHE by $\sin(\pi/6) = 50$ % of the total size expected for the field sweep measurement $|\Delta V_{\text{H}}^{\text{field}}|$, setting the maximum change in the AHE signal.

Significantly, the multi-grain character of Mn$_3$Sn allows us to tune the magnitude of the Hall voltage change by the write current $I_{\text{write}}$ in an analog manner. Figure 3b shows the $V_{\text{H}}$-$I_{\text{write}}$ loops obtained for various values of $I_{\text{write}}^{\text{Min}}$ (Methods). The Mn$_3$Sn/Pt device exhibits a multi-stable signal according to the magnitude of the write current $I_{\text{write}}^{\text{Min}}$. This indicates that our AF device memorizes the amount of electrical current that passes through it. This behavior is known as a memristor and thus would be of great importance for neuromorphic computing[53,62-65].

In recent years, AFMs have attracted significant attention as they have vanishingly small stray fields perturbing neighboring cells and much faster spin dynamics than ferromagnetic counterparts[24-26], leading to the breakthrough of the electrical-current control of AF sublattices and its detection using anisotropic or spin Hall magnetoresistance[54-59]. However, contrary to conventional FM spintronics, these emerging technologies require additional operation schemes to apply current along different directions from the crystalline axes, hampering their integration to conventional spintronics.

In contrast, our experimental demonstration of room-temperature electrical switching of an AF Weyl metal indicates that the topological AFM may replace the active element of a spintronics device. As $Mn_3Sn$ has robust topological properties, its polycrystalline form is practically useful for reading and writing using the same control protocols as those developed for FMs. In addition, the same protocol could be applied not only to $Mn_3Sn$ but also most likely to other members of $Mn_3X$ and more examples of AFMs with a similar magnetic symmetry[51]. A recent report on electrical switching behavior in the antiperovskite $Mn_3GaN$[66] suggests a similar mechanism and further clarification is highly awaited.

Finally, our work using the Weyl AFM $Mn_3Sn$ indicates the SOT switching as a convenient tool to electrically manipulate the distribution of Weyl points and Berry curvature in the momentum space. Thus, our observation of the electrical switching allows significant prospects for developing a new field of topological AF spintronics by studying the dynamics and emergent electromagnetism through integrating topological AFMs with spintronics technologies[33].

63. Kurenkov, A. *et al.* Artificial neuron and synapse realized in an antiferromagnet/ferromagnet heterostructure using dynamics of spin–orbit torque switching. *Adv. Mater.* **31**, 1900636 (2019).

64. Kriegner, D. *et al.* Multiple-stable anisotropic magnetoresistance memory in antiferromagnetic MnTe. *Nat. Commun.* **7**, 11623 (2016).

65. Olejník, K. *et al.* Antiferromagnetic CuMnAs multi-level memory cell with microelectronic compatibility. *Nat. Commun.* **8**, 15434 (2017).

66. Hajiri, T., Ishino, S., Matsuura, K. & Asano, H. Electrical current switching of the noncollinear antiferromagnet $Mn_3GaN$. *Appl. Phys. Lett.* **115**, 052403 (2019).
**Acknowledgements**   We thank Danru Qu, Takahiro Tomita, Yuki Hibino, Takayuki Nozaki, and Shinji Yuasa for useful discussions. We thank Daisuke Nishio-Hamane for the SEM-EDX measurements. This work is partially supported by CREST(JPMJCR18T3), Japan Science and Technology Agency (JST), by Grants-in-Aids for Scientific Research on Innovative Areas (15H05882 and 15H05883) from the Ministry of Education, Culture, Sports, Science, and Technology of Japan, by Grants-in-Aid for Scientific Research (16H06345, 18H03880, 19H00650) and by New Energy and Industrial Technology Development Organization.**Author contributions**   H.T. and T.H. contributed equally to this work. S.N. conceived the project. S.N., K.K., S.M. and Y.O. planned the experiments. T.H., S.M., A.K., Taka.N., and K.Y. prepared and characterized the $Mn_3Sn$ multilayered films. K.K. fabricated the Hall bar devices. H.T., T.H., K.K., and S.M. performed the electrical switching measurements. T.H. performed the magneto-transport measurements and A.S. performed the thermoelectric measurements. Taku.N. and R.A. performed numerical calculations and provided a theoretical explanation. T.H., Taku.N., S.M. and S.N. wrote the manuscript. All authors discussed the results and commented on the manuscript.

**Competing Interests**  The authors declare that they have no competing financial interests.

**Correspondence**  Correspondence and requests for materials should be addressed to S.N. (email: satoru@phys.s.u-tokyo.ac.jp).

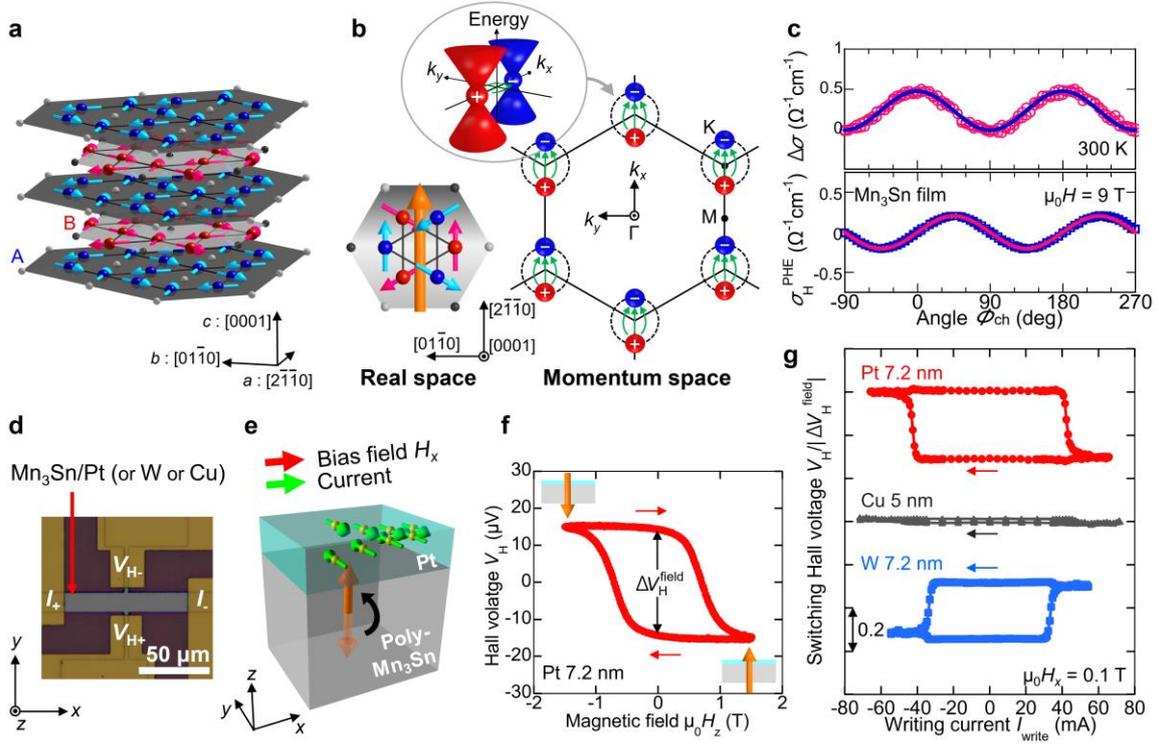

**Figure 1 | Topological Weyl antiferromagnet Mn$_3$Sn and bilayer device layout. a,** Mn$_3$Sn crystal structure and inverse triangular spin (ITS) structure. The large blue and red spheres (small gray and black spheres) represent Mn atoms (Sn atoms) at $z = 0$ and $1/2$, respectively. The Mn magnetic moments (light blue and pink cylindrical arrows) lie within the kagome-layer with the AB-AB stacking sequence and form the ITS structure at room temperature. The spin structure on the kagome bilayers can be viewed as ferroic ordering of a cluster magnetic octupole. **b,** Left: cluster magnetic octupole (orange cylindrical arrow) consisting of the six spins on the kagome bilayer. Right: schematic distribution of the Weyl points near the Fermi energy in the momentum space ($k_x$-$k_y$ plane at $k_z = 0$) for the magnetic structure shown in the left-side figure. Red and blue spheres correspond to Weyl nodes which respectively act as sources (+) and drains (−) of the Berry curvature (green arrows)[15]. Inset: three dimensional schematic picture of a pair of Weyl nodes. **c,** Angular dependence of the longitudinal

magnetoconductivity $\Delta\sigma = \sigma - \sigma_\perp$ and the planar Hall conductivity $\sigma_H^{PHE}$ of the Mn$_3$Sn film at 300 K and 9 T. Blue and pink solid lines for $\Delta\sigma$ and $\sigma_H^{PHE}$ are the fitting results using theoretical equations for the chiral anoamlay, i.e., $\Delta\sigma = \Delta\sigma_{chiral}\cos^2\Phi_{ch}$ and $\sigma_H^{PHE} = \Delta\sigma_{chiral}\sin\Phi_{ch}\cos\Phi_{ch}$, respectively[16,17,21,22]. Here $\Delta\sigma_{chiral}$ (= $\sigma_\parallel - \sigma_\perp$) is the chiral-anomaly induced positive magnetoconductivity, and $\sigma_\parallel$ and $\sigma_\perp$ are the magnetoconductivity when the current is parallel ($\Phi_{ch} = 0°$) and perpendicular ($\Phi_{ch} = 90°$) to the magnetic field, respectively. $\Phi_{ch}$ refers to the in-plane rotation angle between the magnetic field and electrical current directions as presented in Extended Data Fig. 2e. Details can be found in Methods. **d,** Optical micrograph of the fabricated Mn$_3$Sn-nonmagnet bilayer Hall bar devices. Write and read currents are applied along the *x*-direction under the magnetic field $H_x$, $H_y$, or $H_z$ along the *x*, *y*, or *z*-direction, respectively. **e,** Schematic image for the spin-orbit torque switching. The spin polarized current (green cylindrical arrows on yellow spheres) generated in Pt exerts a spin-orbit torque and causes the switching of the polarization axis of the cluster magnetic octupole (orange cylindrical arrow) in the polycrystalline Mn$_3$Sn under a write current and a bias field along the *x*-direction. **f,** Hall voltage $V_H$ vs. magnetic field along the *z*-direction $H_z$ for the Mn$_3$Sn/Pt 7.2 nm device at room temperature. Inset: Schematic figures illustrating the direction of the cluster magnetic octupole. **g,** Hall voltage $V_H$ vs. write current $I_{write}$ for the Pt 7.2 nm, Cu 5 nm, and W 7.2 nm devices at room temperature. The Hall voltage is normalized by the zero-field Hall voltage $|\Delta V_H^{field}|$ in the magnetic field dependence for each sample.

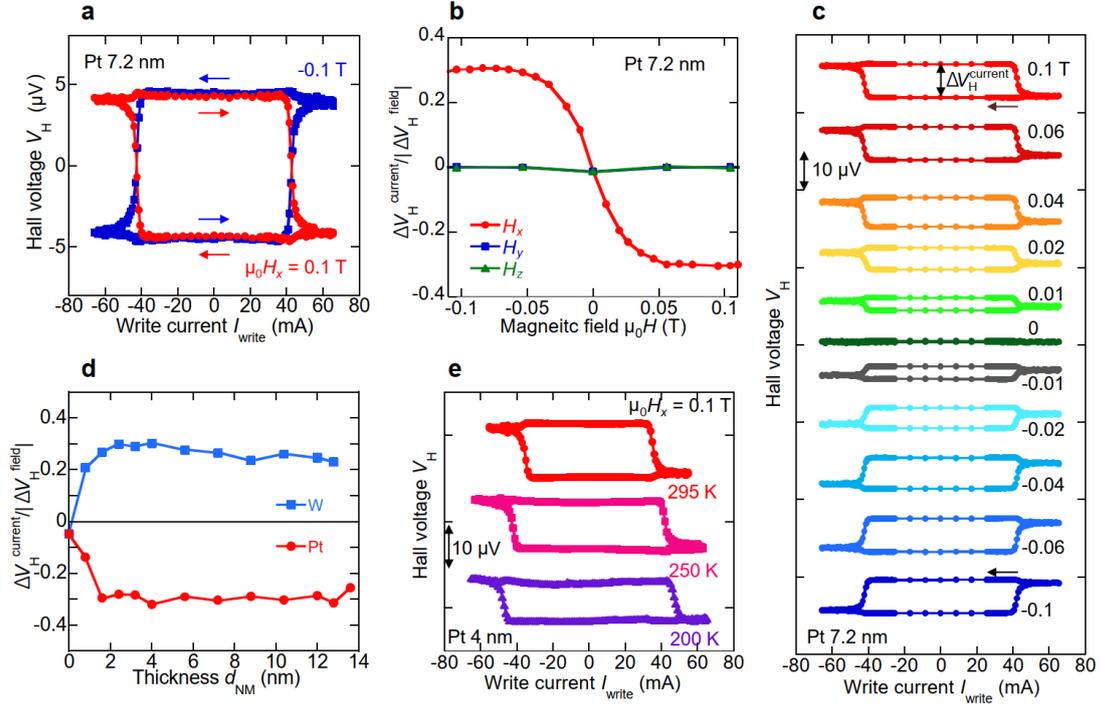

**Figure 2 | Spin-orbit torque induced magnetic switching in the Mn$_3$Sn devices. a,** Write current $I_{write}$ dependence of the Hall voltage $V_H$ ($V_H$ - $I_{write}$ loops) for the Pt 7.2 nm device under the magnetic field along the $x$-direction $H_x$ = +0.1 T (red) and −0.1 T (blue). **b,** Magnetic field dependence of the ratio of the current induced Hall-voltage switching to the field induced Hall-voltage switching $\Delta V_H^{current}/|\Delta V_H^{field}|$ under a field applied along the $x$-, $y$-, and $z$-directions for the Pt 7.2 nm device. **c,** $V_H$ - $I_{write}$ loops for the Pt 7.2 nm device under various $H_x$. **d,** NM-layer thickness dependence of the $\Delta V_H^{current}/|\Delta V_H^{field}|$ under $H_x$ = 0.1 T. Measurements in **a-d** have been performed at room temperature. **e,** $V_H$ - $I_{write}$ loops for the Pt 4 nm device under $H_x$ = 0.1 T at 200, 250, and 295 K.

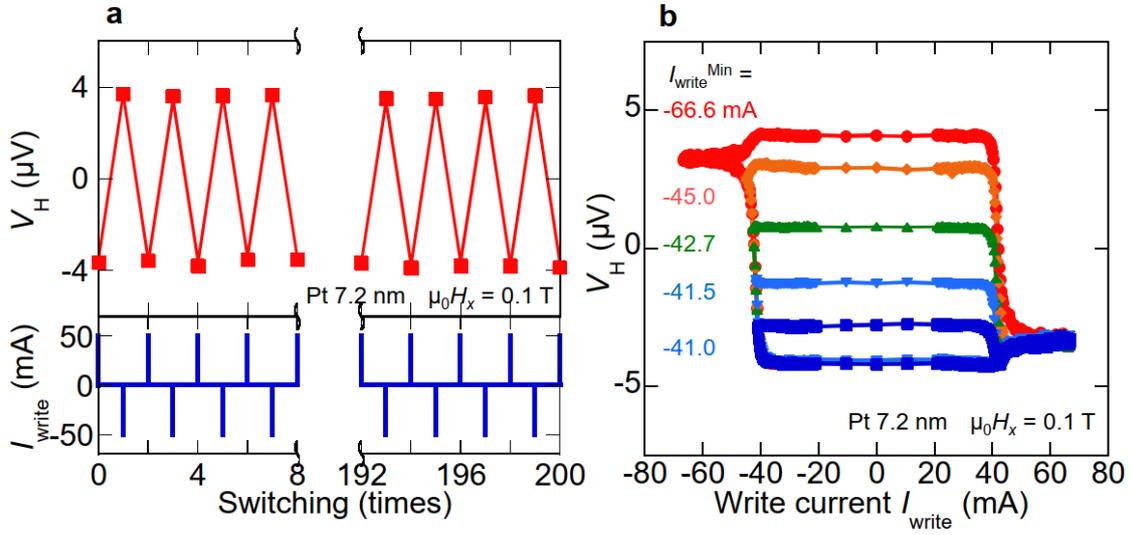

**Figure 3 | Reconfigurable antiferromagnetic switching. a,** Hall voltage $V_H$ (top) and the applied write current $I_{write}$ (bottom) for the Pt 7.2 nm Hall bar devices at room temperature and under the magnetic field along the *x*-direction $H_x$ = 0.1 T in the total 200 times switching of positive and negative $I_{write}$ with a duration of 100 ms. The Hall voltage $V_H$ measured with $I_{read}$ = 0.2 mA changes the sign depending on the polarity of $I_{write}$ = 50 mA. **b,** $V_H$ - $I_{write}$ loops for the Pt 7.2 nm device under $H_x$ = 0.1 T at room temperature. The minimum write current $I_{write}^{Min}$ changes the magnitude of the current induced Hall voltage switching.

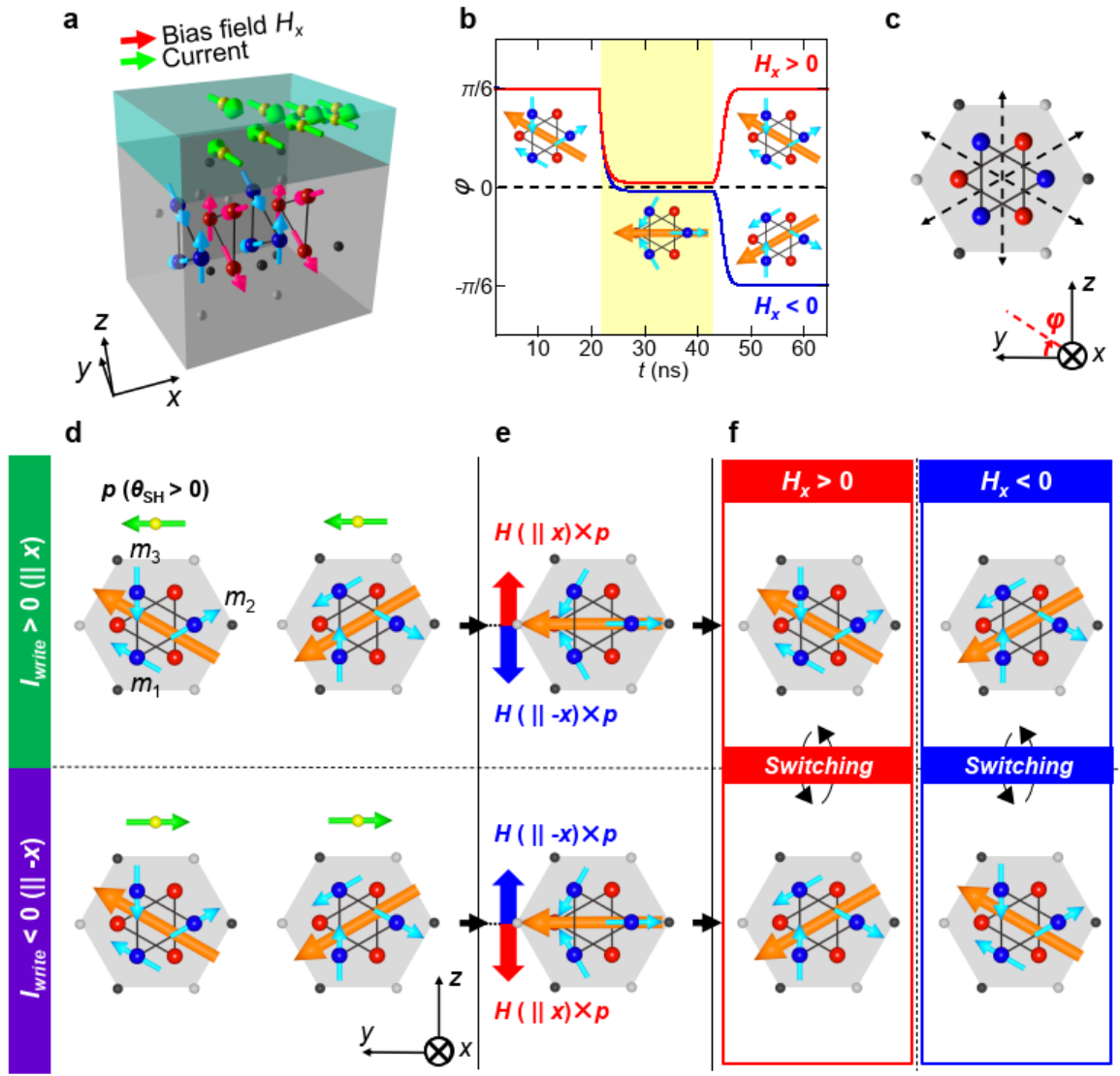

**Figure 4 | Spin-orbit torque mechanism and electrical switching of noncollinear spin texture. a,** Schematic image for the spin-orbit torque switching in the crystal grain configuration (a) under the current and bias field $H$ along the *x*-direction discussed in the main text, where the *c*: (0001)-plane (the kagome-layer) $\perp I$ and the *b*: $[01\bar{1}0]$-direction $\parallel p$. $p$ is parallel to the spin moments of the spin current (green arrows, anti-parallel to the spin angular momentum). **b,** In-plane angle $\varphi$ of the octupole polarization as a function of time *t* for the case where $p \parallel y$ and $H \parallel x$ calculated numerically using the Landau-Lifshitz-Gilbert (LLG) equation discussed in the main text. Pulse current (damping like torque $T_{ia}$) is applied for 21.5 ns $< t <$

42.9 ns (yellow shaded region) and leads to the steady state with the angle $\varphi = 0 \pm \delta$, whose sign is determined by the sign of the bias field $H_x$. **c,** Schematic images of the crystal grains of $Mn_3Sn$ in the configuration (a). Dashed arrows represent the magnetic easy-axis. **d-f,** Switching dynamics of the sublattice moments (light blue cylindrical arrows) and octupole polarization (orange cylindrical arrows) in $Mn_3Sn$ with the configuration (a) presented in Fig. 4a. **d,** Upper (lower) panel corresponds to the initial states before the application of the write current $I_{write}$ along +(−) *x*-direction inducing the spin currents with *p* (green arrow). **e,** Steady states under $I_{write}$. The switching direction of the octupole polarization is determined by $\boldsymbol{H} \times \boldsymbol{p}$ (blue and red arrows). **f,** Final states determined by the sign of *H* (left (red) or right (blue) frame) and the sign of $I_{write}$ (upper or lower panel). The black arrows indicate the switching processes caused by the $I_{write}$ sweep under a positive (left) or negative (right) bias field.

**Methods**

**Sample and device fabrication**

Our samples for the swithching measurements consisting of Ru(2)/Mn$_3$Sn(40)/Pt ($d_{Pt}$ = 0-14), W($d_W$ = 0-14), or Cu($d_{Cu}$ = 0, 5)/AlO$_x$(5) (all thickness in nm) multilayers are grown on thermally oxidized Si wafers (Fig. 1d). Initially, Ru and Mn$_3$Sn are deposited at room temperature using a dc magnetron sputtering machine with a base pressure of less than $5 \times 10^{-7}$ Pa (rate ~0.2 nm/s with 60 W of power and 0.6 Pa Ar gas pressure for the Mn$_3$Sn layer), followed by subsequent annealing in situ at 450 °C for 0.5 h. After cooling to room temperature, Cu, Pt, and/or W layers and AlO$_x$ layer are grown by molecular beam epitaxy (MBE) under ultrahigh vacuum with the base pressure of below $2 \times 10^{-8}$ Pa. We carry out all the fabrication processes in situ conditions including the sample transfer from the sputter chamber to the MBE chamber. The composition of the Mn$_3$Sn layer is Mn$_{3.01(2)}$Sn$_{0.99(2)}$, determined by scanning electron microscopy-energy dispersive X-ray spectrometry (SEM-EDX). Structural analysis is performed by X-ray diffraction (XRD) method using a monochromated CuKα source with a wavelength of 1.54 Å. Mn$_3$Sn is confirmed as a single phase of the D0$_{19}$ structure. For switching measurements, the Mn$_3$Sn multilayers are patterned into Hall bar devices with the in-plane dimensions 16 μm × 96 μm by the standard photolithography combined with a dry etching process and Ti/Au contact pads and wires are deposited by electron beam evaporation as shown in Fig. 1d.

**Transport measurements**

A standard four-probe method with a reading current of $I_{read}$ = 0.2 mA is employed to measure a Hall voltage $V_H$ of the Mn$_3$Sn Hall bar devices as presented in Fig. 1d. In the SOT induced switching measurements, we apply a writing pulse current $I_{write}$ with the duration of 100 ms, and then turn it off. Subsequently, we apply $I_{read}$ and start reading $V_H$ with an interval of 600 ms as shown in Extended Data Fig. 6a. The estimated heating of the central part of the device is ~ 50 K, and the temperature remains lower than the Néel point even when the write current $I_{write}$ is on. Moreover, $V_H$ is not affected by heating effect caused by the $I_{write}$ injection, but represents the value at each monitored temperature since the wait time of 600 ms between the $I_{write}$ injection and $V_H$ measurement (Extended Data Fig. 6a) is long enough, as confirmed in the wait time dependence of $V_H$ (Extended Data Fig. 7d). Except for the low-temperature results presented in Fig. 2e, measurements are all performed at room temperature (~ 293 K). To estimate the current density in the non-magnetic metal (NM), the resistivity of Ru/Mn$_3$Sn, Pt and W are calculated based on the Si/SiO$_2$/Ru(2)/Mn$_3$Sn(40)/AlO$_x$(5) and Si/SiO$_2$/Ru(2)/Mn$_3$Sn(40)/Pt or W($d_{NM}$)/AlO$_x$(5) Hall bar devices by using the conductors-in-parallel model. A two-probe method with the contact pads for the current is used for this resistivity measurement. The thickness dependence of the resistivity in the NM layer is shown in Extended Data Fig. 6b. The resistivity of the Mn$_3$Sn(40) thin films deposited by the same method as the one used in our present work is ~300 μΩcm[27], consistent with the measurement results for a bulk[10]. The electrical measurement presented in Fig. 2e is performed under 6 x 10$^{-5}$ Pa within a vacuum chamber cooled by a Helium compressor. The sample substrate is fixed on a Cu sample stage and the temperature is measured by a thermometer inside the sample stage. In the $V_H$- $I_{write}$ loops measurements presented in the Fig. 3b, we first initialize the magnetic domains by applying a positive pulse current of $I_{ini}$ larger than the switching current $I_c$. Then, we scan the current $I_{write}$ ranging from $I_{ini}$ to $I_{write}^{Min}$ under $H_x$ = +0.1 T.

For the longitudinal magnetoconductance (MC) and planar Hall effect (PHE) measurements, the Mn$_3$Sn 40 nm layer deposited directly on the Si/SiO$_2$ substrates is employed, which is fabricated using the same synthesis condition as the multi-layer for the switching measurements. As shown in Extended Data Figs. 1e and 2e, the bar-shaped samples with the voltage terminals made by the Ag paste are measured. To obtain the magnetic-field angular dependence, the horizontal rotator option of the commercial physical property measurement system (PPMS) is used. Similar to the MR and PHE measruements, the bar-shaped samples (9 mm-length × 2 mm-width) are used in the commercial system PPMS to measure the transport properties including the resistivity, Hall, Seebeck, and Nernst effects presented in Extended Data Fig. 4. The electrical and thermal currents are applied along the length-direction, and the magnetic field is applied along the perpendicular direction to the film plane. The transverse thermoelectric conductivity $\alpha_N$ is estimated from the electrical conductivity $\sigma$, Hall conductivity $\sigma_H$, Seebeck coefficient $S_{SE}$ and Nernst coefficient $S_{ANE}$ using the equation $\alpha_N = \sigma S_{ANE} + \sigma_H S_{SE}$[38].

**Experimental evidence for the Weyl metal state in the polycrystalline Mn$_3$Sn thin film.**

To demonstrate the switching of the topological antiferromagnetic state, we first need to show that the polycrystalline sample does support the same Weyl semimetal state as has been found in Mn$_3$Sn single crystals[13,15,28].

In general, the evidence for Weyl fermions can be obtained by various probes such as,
1)  Angle resolved photoemission spectroscopy (ARPES) to reveal the band structure,
2)  Chiral anomaly through magnetoconductance and planar Hall effect.

In addition, when the Weyl fermions are due to time-reversal symmetry breaking,

3) the observation of the large anomalous Nernst effect (ANE) beyond the empirical scaling law with magnetization is known to provide further strong evidence for Weyl fermions[13,14,18,23].

In the bulk single crystal case of Mn$_3$Sn, all three probes above have been used to establish the magnetic Weyl semimetal state[13,15,28]. To further obtain experimental evidence for the switching of the topological antiferromagnet and manipulation of the distribution of the Weyl points in our polycrystalline thin films, we employ the following approach using the transport probes (2) and (3) for the Weyl fermions, as both probes are also useful for polycrystalline samples. First, the observation of the chiral anomaly should provide strong evidence for Weyl fermions (see e.g. Refs [1,4,6,7,9,12,15-17,21,22]). Second, if one could detect the large ANE much larger than what would be expected based on the empirical scaling law with magnetization, it should provide another strong evidence for the Weyl points near the Fermi energy, which produces the large Berry curvature[13,14,18,23]. Third, since ANE probes not only the size but also the sign of the Berry curvature at the Fermi energy[38], the motion of the Weyl points producing the Berry curvature should be detected as the sign change in ANE. Thus, the combination of the chiral anomaly and the sign change in the large ANE beyond the magnetization scaling should provide clear evidence for the motion of the Weyl points near the Fermi energy in the momentum space.

**I. Chiral Anomaly**

A. Positive longitudinal magnetoconductivity (MC)

As we mentioned, the chiral anomaly has been established as the evidence for Weyl fermions (see e.g. Refs [1,4,6,7,9,12,15-17,21,22]). It refers to the positive longitudinal MC due to the violation of a separate number conservation law of the three-dimensional left- and right-handed Weyl fermions. The anomaly is known to arise when the magnetic field ***H*** is aligned parallel to the electric field ***E***. As this effect basically depends on the scalar product |***E·H***|, the qualitative feature should mostly depend on the relative angle between ***E*** and ***H*** and should

appear in a polycrystalline sample. Thus, if the Weyl fermions are present in the polycrystalline form of Mn$_3$Sn, the chiral anomaly should lead to an anisotropic MC, namely, the *positive longitudinal* MC when the magnetic field is applied parallel to the electric field ($H \parallel E$), while the *negative transverse* MC should appear when the magnetic field is applied along the perpendicular direction to the electrical current ($H \perp E$).

To examine the possibility of the chiral anomaly in our polycrystalline films, we have carried out the magneto-transport measurements. We find that the anisotropic field dependence of the MC is fully consistent with the chiral anomaly as shown in Extended Data Fig. 1a. Namely, the longitudinal MC for the magnetic field parallel to the current, $H \parallel E$ ($\Theta_{ch} = 0°$), increases with $|H|$. On the other hand, the field perpendicular to the electric field, $H \perp E$ ($\Theta_{ch} = 90°$) generates much smaller but a finite increase with $|H|$, which is attributable to the suppression of the spin fluctuations. Generally, in magnetic conductors, the application of the magnetic field may lead to the suppression of thermally induced spin fluctuations, and thus the positive MC, irrespective of the angle $\Theta_{ch}$ between $H$ and $E$. Although this effect may complicate the detection of the chiral anomaly, this type of the positive longitudinal MC should become reduced on cooling. In sharp contrast, we find that the longitudinal MC for $H \parallel E$ becomes enhanced by nearly 50 % by decreasing the temperature from 300 K to 250 K (Extended Data Figs. 1a-1d). This indicates that the longitudinal MC does not come from the suppression of the spin fluctuations, but the chiral anomaly itself. In addition, the transverse MC for $H \perp E$ becomes negative under the fields higher than 5 T (Extended Data Fig. 1b), indicating the effect of the spin fluctuations can be suppressed under high fields at 250 K. Thus, our experiment indicates that the major origin of the anisotropic MC is the chiral anomaly both at 250 K and 300 K.

Another conventional mechanism that may yield the positive longitudinal MC is the current-jetting effect. In high mobility semimetals, the orbital effect leads to a large negative transverse MC in comparison with the longitudinal MC[4,67-69]. This anisotropic MC may orient the current to the direction parallel to the magnetic field and suppress the transverse component of the current flow, resulting in the positive longitudinal MC that strongly depends on the sample geometry. Recent studies have revealed that the positive longitudinal MC observed in several Dirac and Weyl semimetals comes from the current-jetting effect[4,69]. This effect is unlikely to occur in Mn$_3$Sn as both the mobility and the anisotropy in the MC are much lower than those reported for the above semimetals. Nevertheless, we have examined the possible current inhomogeneity by preparing three sets of pairs of the voltage terminals placed on the centerline as well as at both sides of a Mn$_3$Sn polycrystalline film (Extended Data Fig. 1e). All three voltage probes detect the same field dependence of the longitudinal MC and transverse MC, confirming that the current distribution is homogeneous inside the film and ruling out the current jetting effect as the origin of the positive longitudinal MC (Extended Data Fig. 1a).

B. Planar Hall Effect

Recent theory has shown that the chiral anomaly may lead to not only the anisotropic MC (positive longitudinal MC) but the planar Hall effect (PHE)[16,17,21,22]. The conductivity $\sigma$ and planar Hall conductivity $\sigma_H^{PHE}$ are respectively formulated as,

$$\sigma = \sigma_\perp + \Delta\sigma_{chiral} \cos^2\Phi_{ch} \quad (2)$$

$$\sigma_H^{PHE} = \Delta\sigma_{chiral} \sin\Phi_{ch}\cos\Phi_{ch} \quad (3)$$

Here, $\Delta\sigma_{chiral}$ ($= \sigma_\parallel - \sigma_\perp$) is the positive magnetoconductivity induced by the chiral anomaly, where $\sigma_\parallel$ and $\sigma_\perp$ are the magnetoconductivity when the electrical current is parallel ($\Phi_{ch} = 0°$) and perpendicular ($\Phi_{ch} = 90°$) to the magnetic field, respectively. $\Phi_{ch}$ refers to the in-plane angle between the magnetic field and electrical current directions as presented in Extended Data Fig. 2e.

Extended Data Figs. 2a,b and 2c,d present the in-plane rotation angle ($\Phi_{ch}$) dependence of the MC $\Delta\sigma = \sigma - \sigma_\perp$ and the planar Hall conductivity $\sigma_H^{PHE}$, respectively. As we noted above, the electrical current direction as well as the sample is rotated in the plane with the angle ($\Phi_{ch}$) from the magnetic field direction (Extended Data Fig. 2e) The anisotropic MC $\Delta\sigma$ exhibits the sinusoidal dependence on the rotation angle ($\Phi_{ch}$). Our results obtained at 300 K and 250 K for both anisotropic MC and PHE are fully consistent with theory, and well fit by the hypothetical formula described in Eqs. (2) and (3). Further consistent with the theory described in Eqs. (2) and (3), the same $|\Delta\sigma_{chiral}|$ value ~0.4 $\Omega^{-1}$cm$^{-1}$ (300 K) and ~1 $\Omega^{-1}$cm$^{-1}$ (250 K) are obtained for both $\Delta\sigma(\Phi_{ch})$ and $\sigma_H^{PHE}(\Phi_{ch})$.

To summarize, our measurements of both magnetoconductance and the PHE provide strong evidence for the chiral anomaly and thus for the Weyl fermions in the polycrystalline thin films of Mn$_3$Sn.

**II. Large anomalous Nernst effect (ANE) beyond the empirical scaling law with magnetization**

Further evidence for the Weyl fermions can be obtained from the measurements of ANE. The ANE is usually found in ferromagnets and empirically known to scale with magnetization[13]. In magnetic Weyl semimetals, however, the Weyl points near the Fermi energy should generate a large net Berry curvature, leading to a large ANE beyond the empirical scaling law with magnetization[13,14,18,23]. A prominent example has been found in the case of the Mn$_3$Sn single crystals, as shown in Extended Data Fig. 3. Namely, despite its vanishingly small magnetization ~ 3 m$\mu_B$/Mn, Mn$_3$Sn exhibits the large ANE (the anomalous Nernst coefficient $S_{ANE}$ ~ 0.35 µV/K), equivalent to those obtained for ferromagnets with a large magnetization of ~1 µ$_B$ according to the empirical scaling law (Extended Data Fig. 3) [13].

To confirm the large ANE driven by the Weyl fermions, we have newly measured the ANE of our polycrystalline Mn$_3$Sn thin film. The field dependence of the anomalous Nernst coefficient $S_{ANE}$ and Hall resistivity $\rho_H$ are presented in Extended Data Fig. 4a. The polycrystalline Mn$_3$Sn film exhibits a large spontaneous ANE of $S_{ANE}$ ~ 0.25 µV/K at 300 K similar to the single crystalline case. On the other hand, the spontaneous magnetization of the film is found to be only ~ 6 m$\mu_B$/Mn[27], again similar to the case for the single crystals. Thus, $S_{ANE}$ obtained for the film is more than two orders of magnitude larger than the value expected from the empirical scaling law with magnetization (Extended Data Fig. 3).

Generally, ANE is made of two components, $S_{ANE} = \alpha_N/\sigma - S_{SE}(\sigma_H/\sigma)$, where $\alpha_N$ is the transverse thermoelectric coefficient, $\sigma$ the electrical conductivity, and $S_{SE}$ the Seebeck coefficient. If ANE is enhanced due to the Weyl points near the Fermi energy $E_F$, the first term should be dominant. This is because the Berry curvature $\Omega_{n,z}(\boldsymbol{k})$ near $E_F$, which is exactly what is significantly enhanced by the Weyl points near $E_F$, determines the thermoelectric coefficient $\alpha_N$, as described in the following,

$$\alpha_N \sim \sum_{n,\boldsymbol{k}} \Omega_{n,z}(\boldsymbol{k})\delta(E_F - \varepsilon_{n,\boldsymbol{k}}) \quad (4)$$

while the anomalous Hall conductivity measures the sum of the Berry curvature of all the occupied bands[38].

To confirm this, we made the analysis and found that the transverse thermoelectric coefficient, $\alpha_N = \sigma S_{ANE} + \sigma_H S_{SE} \sim 0.1$ AK$^{-1}$m$^{-1}$ at 300 K (Extended Data Fig. 4b), is ~ 60% of the value obtained for the bulk single crystals[13] (See Methods and the caption of Extended Data Fig. 4 for the estimate of $\alpha_N$ using the results of the resistivity, ANE, AHE, and Seebeck effect measurements.). Note that the first term $\sigma S_{ANE} \sim 0.09$ AK$^{-1}$m$^{-1}$ is much larger than the second $\sigma_H S_{SE} \sim 0.01$ AK$^{-1}$m$^{-1}$, namely, $\alpha_N \sim \sigma S_{ANE}$, indicating that the large $\alpha_N$ due to the large Berry curvature near $E_F$ dominates $S_{ANE}$. The large spontaneous value of $S_{ANE}$ and $\alpha_N$ provides further strong evidence for the Weyl fermions in the Mn$_3$Sn polycrystalline film.

**III. Sign change in ANE as evidence for the switching of the topological antiferromagnet**

In a magnetic Weyl semimetal, the breaking of the time-reversal symmetry (TRS) generates a pair of Weyl points in the momentum space, and the symmetry constraint requires the nodal direction connecting a pair of the Weyl points to be aligned along the TRS breaking field or order parameter. Thus, reversing the magnetic field direction or order parameter rotates the pairs of the Weyl points by 180º in the momentum space. In Mn$_3$Sn, the reversal of the cluster octupole polarization flips the pairs of the Weyl points and the associated Berry curvature, which can be viewed as an axial vector[15], as illustrated in Extended Data Fig. 5.

Equation (4) indicates that such rotation of the Berry curvature should lead to the sign change in $\alpha_N$. As discussed above, $\alpha_N$ for the polycrystalline Mn$_3$Sn thin film is highly enhanced owing to the Weyl points near the Fermi energy as in the case of the single crystals. In such a case of a magnetic Weyl semimetal, therefore, the sign change in $\alpha_N$ observed in the field sweep should come from the 180º rotation of the pairs of the Weyl points near $E_F$. Moreover, as shown in Extended Data Fig. 4b, $\alpha_N$ and $\sigma_H$ show the sign change by reversing the magnetic field, and their field dependences overlap on top of each other in the thin film case, further indicating that the sign change in $\sigma_H$ should also come from the flipping of the pairs of the Weyl points. Thus, we conclude that the AHE signal ($V_H$) jump induced by the electrical switching probes the rotation of the pairs of the Weyl points and the associated Berry curvature in the Mn$_3$Sn thin films. In particular, for the case of the model described in Fig. 4, the switching should take place between the two different configurations of Weyl points shown as Panels (**a**) and (**d**) in Extended Data Fig. 5.

In summary, all our measurements and analyses discussed above demonstrate that the same Weyl physics as found in the bulk Mn$_3$Sn is applicable to the case for the polycrystalline thin films as well. Moreover, highly enhanced $\alpha_N$ due to the large Berry curvature near Fermi energy shows the sign change by reversing the magnetic field, indicating the 180º rotation of the pairs of the Weyl points. Thus, the electrical switching of the AHE signal in the Mn$_3$Sn thin film indicates the electrical manipulation of the topological antiferromagnetic state and the distribution of Weyl points in the momentum space.

**Landau-Lifshitz-Gilbert (LLG) equation based on the one-layer kagome lattice Hamiltonian**

To understand the deterministic switching mechanism based on SOT, we study the dynamics of the sublattice moments $\boldsymbol{m}_{ia}$ obeying the Landau-Lifshitz-Gilbert (LLG) equation[29,60] based on the following Hamiltonian defined on the one-layer kagome lattice (one blue layer on the $yz$-plane in Fig. 4a);

$$\mathcal{H} = \mathcal{J} \sum_{\langle ia,jb \rangle} \boldsymbol{m}_{ia} \cdot \boldsymbol{m}_{jb} + \mathcal{D} \sum_{\langle i,j \rangle} \boldsymbol{x} \cdot \left( \boldsymbol{m}_{i1} \times \boldsymbol{m}_{j2} + \boldsymbol{m}_{i2} \times \boldsymbol{m}_{j3} + \boldsymbol{m}_{i3} \times \boldsymbol{m}_{j1} \right) - \mathcal{K} \sum_{ia} \left( \boldsymbol{k}_a \cdot \boldsymbol{m}_{ia} \right)^2 \quad (5)$$

where the suffixes *i* and *j* denote a unit cell, *a* and *b* = (1, 2, 3) a sublattice, and $\boldsymbol{x} = (1,0,0)$ the unit vector along *x*-axis (Fig. 4a in the main text). Note that the present model only includes three Mn atoms in a unit cell, and we neglect inter-layer couplings because they do not give qualitatively new effects[48]. Each $\boldsymbol{m}_{ia}$ denotes a unit magnetic moment. The nearest-neighbor exchange interaction $\mathcal{J}$, Dzyaloshinskii-Moriya (DM) interaction $\mathcal{D}$, and in-plane magnetic anisotropy $\mathcal{K}$ are assumed to be positive, and stabilizes the inverse triangular spin (ITS) texture as observed in Mn$_3$Sn[50,70]. The $\boldsymbol{k}_a = (0, \cos\varphi_a, \sin\varphi_a)$ with $(\varphi_1, \varphi_2, \varphi_3) = (11\pi, 3\pi, 7\pi)/6$ lifts the in-plane U(1) degeneracy and fixes the six-fold symmetry. Here, $\mathcal{J}$ and $\mathcal{D}$ are identical to $JS^2$ and $DS^2$ employed in Refs. [61] and [71], and a different definition is adapted for the in-plane anisotropy $\mathcal{K}$ leading to the six-fold magnetocrystalline anisotropy and spontaneous magnetization. The bias field is included in a Zeeman coupling term, $H_{\text{Zeeman}} = -\mu_0 M_s \sum_{ia} \boldsymbol{m}_{ia} \cdot \boldsymbol{H}$, where $M_S = 3$ μ$_B$ is a saturation magnetic moment of Mn atom, and SOT causes the spin-transfer-induced in-plane torque, $\boldsymbol{T}_{ia} = \frac{\hbar |\gamma| J_0 \theta_{\text{SH}}}{2 e m_s d} \boldsymbol{m}_{ia} \times (\boldsymbol{y} \times \boldsymbol{m}_{ia})$. Here, $\theta_{\text{SH}}$ denotes the spin Hall angle in Pt[29], *d* the thickness of Mn$_3$Sn layer, $m_S = 6M_S$ /unit cell volume, $J_0$ the current density, and $\boldsymbol{y} = (0,1,0)$ the unit vector along the *y*-axis (Fig. 4a in the main text). Based on them, the LLG equation becomes equation (1) in the main text [29,60],

$$\dot{\boldsymbol{m}}_{ia} = -|\gamma| \boldsymbol{m}_{ia} \times \boldsymbol{H}_{\text{eff},ia} + \alpha \boldsymbol{m}_{ia} \times \dot{\boldsymbol{m}}_{ia} + \boldsymbol{T}_{ia}.$$

Here, the effective magnetic field $\boldsymbol{H}_{\text{eff},ia}$ is defined as $\boldsymbol{H}_{\text{eff},ia} = -M_S^{-1} \delta\mathcal{H}/\delta\boldsymbol{m}_{ia}$. The first, second, and third terms of r.h.s. represent respectively gyroscopic torque, Gilbert-damping torque, and external torque due to the spin injection. $\gamma$ (< 0) represents the gyromagnetic ratio of an electron, and *α* a Gilbert-damping coefficient.

For the simulation of the magnetization switching process in the configuration (a), we set parameters $J = 23$ meV, $\mathcal{D} = 1.6$ meV, $\mathcal{K} = 0.17$ meV and $\alpha = 0.003$. These are consistent with the experimentally obtained physical parameters in Mn$_3$Sn, where six-fold magnetocrystalline anisotropy energy and spontaneous magnetization are 310 J/m and 0.01 $\mu_B$/f.u., respectively[61]. The other parameters are set by: $\theta_{SH} = 0.1$, $d = 40$ nm, $\mu_0|H| = 0.1$ T and $J_0 = 6 \times 10^{14}$ A/m$^2$.

**Three crystal grain configurations in the polycrystalline Mn$_3$Sn Hall bar devices for the SOT induced switching**

When we consider the possible SOT mechanism for the non-collinear AF spin texture in Mn$_3$Sn using the accumulated spin at the Mn$_3$Sn/NM interface, the polycrystalline character of our Mn$_3$Sn layer suggests the three crystal grain configurations representing the Hall bar devices: (a) the kagome-layer $\perp$ the current $I$ and $\parallel$ the electrically injected carrier spin polarization $p$ (Fig. 4a, Extended Data Fig. 8a), (b) the kagome-layer $\parallel I$ and $\perp p$ (Extended Data Fig. 8b), and (c) the kagome-layer $\parallel I$ and $\parallel p$ (Extended Data Fig. 8c).

As we discuss in the main text, in the configuration (a), the combination of the spin current and the bias field may cause the deterministic SOT switching (Fig. 4, Extended Data Fig. 8a). On the other hand, for the grains with the configuration (b), i.e. the kagome-layer $\parallel I$ and $\perp p$ (Extended Data Fig. 8b), the spin current injection causes a steady state with continuous rotation of a slightly canted ITS structure, destabilizing exchange interaction and DM interaction energy. Extended Data Figure 8d and its inset indicate an example of numerical simulation where the spin injection starts at a time $t = 21.5$ ns. There, we find that the in-plane angle $\varphi$ (as defined in Fig. 4c) oscillates at $t > 21.5$ ns with the frequency of ~3.5 THz as a result of the continuous rotation. This indicates that a random configuration of the octupole moments arises in each grain, depending on the timing when the current is turned off[71,72] and thus the final state is not uniquely determined by the bias field. In the configuration (c) (Extended Data Fig. 8c), such a grain does not have any out-of-plane ($z$) component of the octupole moment and thus does not exhibit any AHE.

Note that the configuration (a) (the kagome-layer $\perp I$ and $\parallel p$) has two characteristic arrangements of the kagome-layer orientations: (a-1) $b$-axis:$[01\bar{1}0] \parallel y$ (used in the theoretical calculation, see Fig. 4a) and (a-2) $a$-axis:$[2\bar{1}\bar{1}0] \parallel y$, as presented in Extended Data Figs. 9a and 9b, respectively. In the field-induced switching, both arrangements should contribute to AHE. In the electrical switching, however, only the arrangement (a-1) in Extended Data Fig. 9a may contribute to AHE since the octupole polarization (orange arrows) in the arrangement (a-2) in Extended Data Fig. 9b is pinned along its easy-axis ($a$-axis:$[2\bar{1}\bar{1}0]$). Thus, a part of the grains with the configuration (a) may contribute to the electrical switching.

**Deterministic switching in numerical simulations**

Here, we discuss the simulation results on the deterministic switching. As is mentioned in the main text, irrespective of the sign of $I_{\text{write}}$, we observe the same steady states in the switching process. This behavior is counterintuitive, as it is contrary to the analogy of conventional ferromagnets, where the magnetic moments are aligned along the direction of $T_{ia}$, which depends on the sign of $I_{\text{write}}$. Here, we show that the damping torque combined with SOT, $T'_{ia} = \alpha m_{ia} \times T_{ia}$, is the key to obtain this behavior.

Extended Data Fig. 10a shows the evolution of the in-plane motion of the octupole polarization without the bias field, where the onset of the pulse current injection is set at $t_1$. While both positive and negative $I_{\text{write}}$ lead to the same steady states after long enough time, we note that in the first short period right after the onset time $t_1$, the octupole polarization actually rotates to the directions determined by $T_{ia}$ depending on the sign of $I_{\text{write}}$ (Inset of Extended Data Fig. 10a). In particular, for the $I_{\text{write}} < 0$ case, we find that the in-plane angle $\varphi$ returns to the initial value at $t_2$ ($> t_1$). Here, we stress that the in-plane components of each sublattice moment at $t_1$ and $t_2$ are almost identical in this $I_{\text{write}} < 0$ case. However, the out-of-plane components of those, which are induced by $T'_{ia} \sim \alpha m_{ia} \times T_{ia}$, are different, as is depicted in Extended Data Fig. 10b. This result indicates that a type of non-adiabatic torque emerges from $T'_{ia}$, which acts on the spins independently of the sign of $I_{\text{write}}$. Namely, the amplitude of this $I_{\text{write}}$-sign independent torque gradually increases with increasing the staggered out-of-plane components induced by $T'_{ia}$, and finally overcomes the $I_{\text{write}}$-sign dependent $T_{ia}$ for $t > t_2$. Note that a usual low-energy effective model, as is discussed in Refs. [48] and [71], cannot capture the above feature since it only includes the in-plane motion of the order parameter.

Although the microscopic switching mechanism is different, we can employ for $Mn_3Sn$ the same conventional spintronics setup as is used in the ferromagnetic devices since the switching conditions are the same. This is achieved with the help of the Gilbert damping torque combined with SOT and the staggered out-of-plane components of sublattice moments.

**Data availability**

The data that support the findings of this study are available from the corresponding author upon reasonable request.

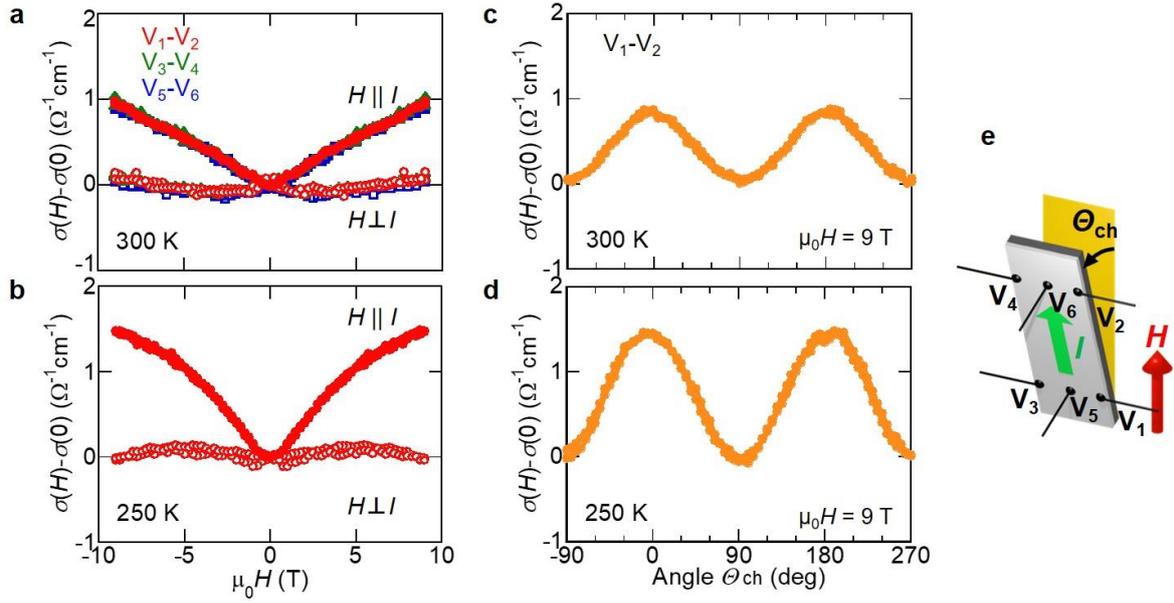

**Extended Data Figure 1 | Longitudinal magnetoconductivity measurements for the Mn₃Sn thin film. a,b,** Magnetic field $H$ dependence of the magnetoconductivity $\sigma(H) - \sigma(0)$ of the Mn₃Sn 40 nm film under $H \parallel$ the current $I$ and under $H \perp I$ at 300 K (**a**) and 250 K (**b**). Here $\sigma(0)$ is the magnetoconductivity at 0 T. **c,d,** Angular $\Theta_{ch}$ dependence of the magnetoconductivity $\sigma(H) - \sigma(0)$ of the Mn₃Sn 40 nm film at 300 K (**c**) and 250 K (**d**). **e,** Schematic figure for the experimental set up for the magneto-transport measurements. To examine the current homogeneity, we employ three pairs of voltage terminals, namely, $V_1$-$V_2$ and $V_3$-$V_4$ at both sides and $V_5$-$V_6$ on the centerline of the film in the measurement for (**a**). Except for (**a**), the pair of the terminals $V_1$-$V_2$ is used.

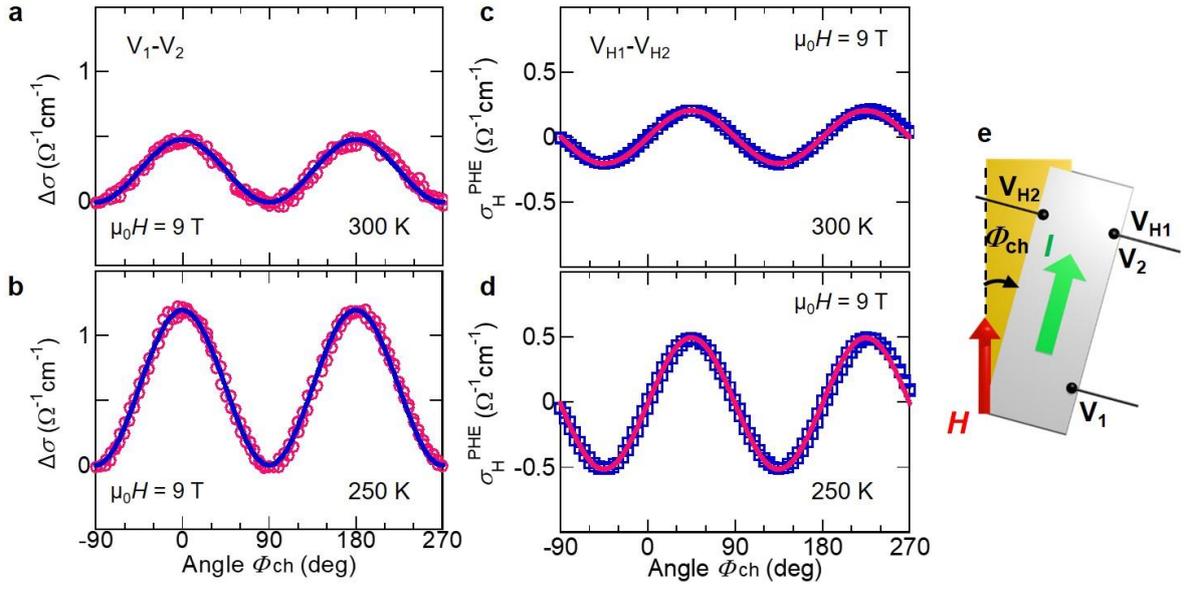

**Extended Data Figure 2 | Longitudinal magnetoconductivity and planar Hall conductivity measurements for the Mn$_3$Sn thin film. a,b,** Angular $\Phi_{ch}$ dependence of the longitudinal magnetoconductivity $\Delta\sigma = \sigma - \sigma_\perp$ of the Mn$_3$Sn 40 nm film at 300 K (**a**) and 250 K (**b**). A blue solid line is the fitting results using Eq. (2). **c,d,** Angular $\Phi_{ch}$ dependence of the planar Hall conductivity $\sigma_H^{PHE}$ of the Mn$_3$Sn 40 nm film at 300 K (**c**) and 250 K (**d**). A pink solid line is the fitting results using Eq. (3). **e,** Schematic figure for the measurement set up for the longitudinal magnetoconductivity and planar Hall effect with the pairs of the terminals V$_1$-V$_2$ and V$_{H1}$-V$_{H2}$, respectively. $\Phi_{ch}$ refers to the in-plane angle between the magnetic field $H$ (red arrow and broken line) and electrical current $I$ (green arrow) directions. Extended Data Figs. 2a and 2c are displayed in the main text as Fig. 1c.
41

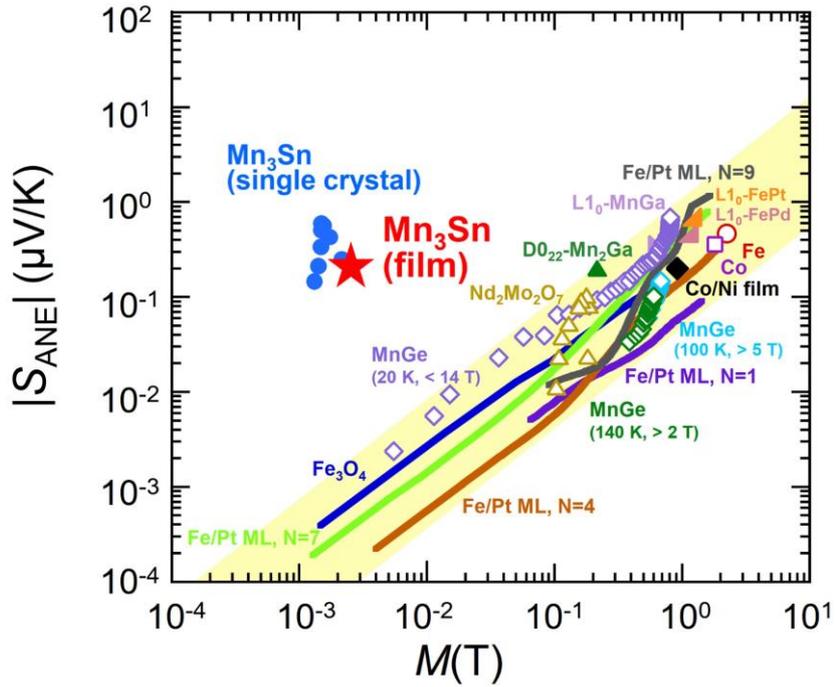

**Extended Data Figure 3 | Large anomalous Nernst effect (ANE) in the $Mn_3Sn$ thin film beyond the empirical scaling with magnetization.** Full logarithmic plot of the anomalous Nernst coefficient $|S_{ANE}|$ vs. magnetization $M$ for a variety of the ferromagnetic metals (yellow shaded region highlighting the empirical scaling law with $M$), for $Mn_3Sn$ single crystal at various temperature including 300 K (blue circle), and for the polycrystalline $Mn_3Sn$ thin film at 300 K (red star) used in the present study. Data is adapted from Ref. [13].

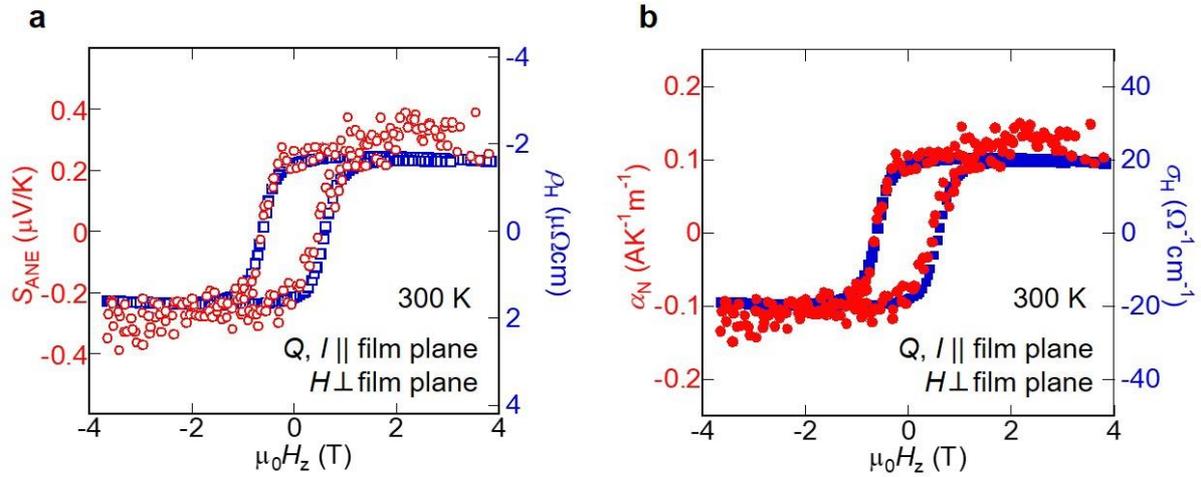

**Extended Data Figure 4 | Field-induced sign change in transverse thermoelectric conductivity and Hall conductivity of the Mn$_3$Sn thin film. a,b,** Field dependence of the anomalous Nernst coefficient $S_{ANE}$ and the Hall resistivity $\rho_H$ (**a**) and the transverse thermoelectric conductivity $\alpha_N$ and the Hall conductivity $\sigma_H$ (**b**) of the Mn$_3$Sn 40 nm thin film at room temperature. The Seebeck coefficient $S_{SE}$ and the resistivity $\rho$ are also measured at 300 K and found to be constant ($S_{SE}$ = 7.6 µV/K and $\rho$ = 290 µΩcm) in the field sweep measurements below ±4 T. $\alpha_N$ is estimated from the electrical conductivity $\sigma = 1/\rho$, the Hall conductivity $\sigma_H = -\rho_H/\rho^2$, the Seebeck coefficient $S_{SE}$ and the Nernst coefficient $S_{ANE}$ by using the equation $\alpha_N = \sigma S_{ANE} + \sigma_H S_{SE}$ (Methods). Here, the heat current $Q$ and electrical current $I$ are applied parallel to the film plane, while the field is along the normal direction ($z$-direction) to the film plane.
43

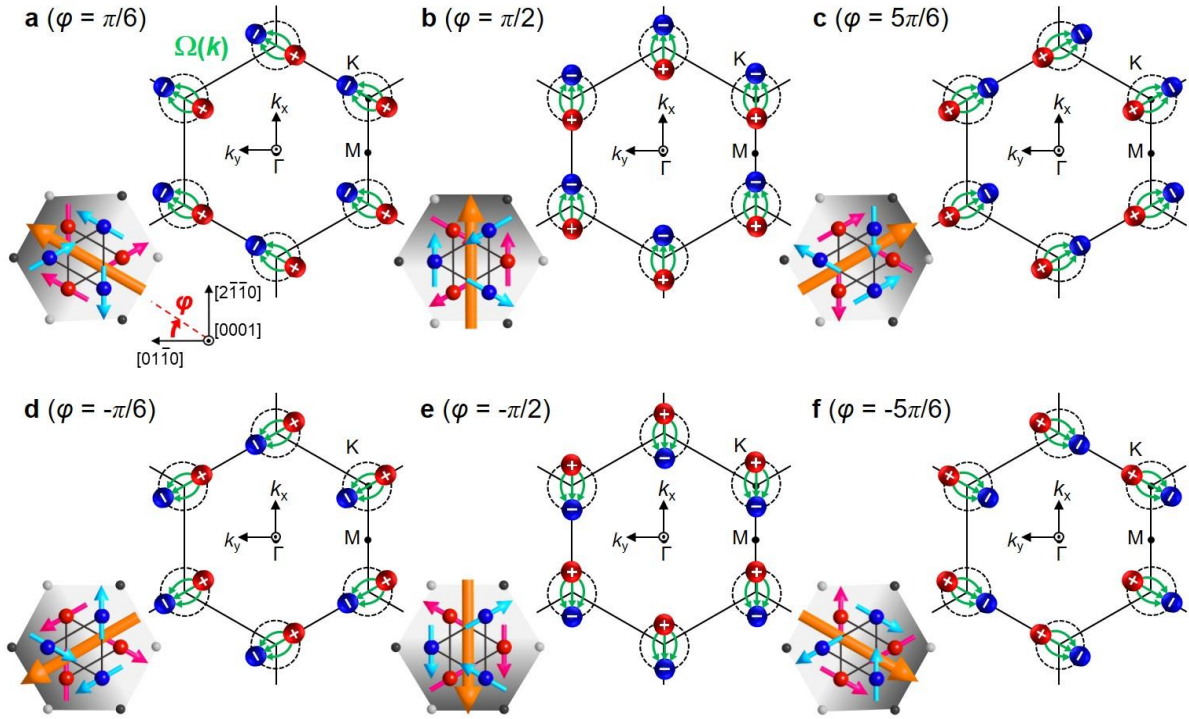

**Extended Data Figure 5 | Control of the nodal direction connecting a pair of Weyl points by the magnetic octupole polarization.** Schematic figures of a cluster magnetic octupole (orange cylindrical arrow) consisting of the six spins on the kagome bilayer in real space (left side figure in each panel) and schematic distributions of the Weyl points near the Fermi energy in the momentum space ($k_x$-$k_y$ plane at $k_z = 0$, right side figure in each panel) for each magnetic structure of $Mn_3Sn$ corresponding to $\varphi = \pi/6$ (**a**), $\pi/2$ (**b**), $5\pi/6$ (**c**), $-\pi/6$ (**d**), $-\pi/2$ (**e**), $-5\pi/6$ (**f**). Red and blue spheres represent Weyl nodes which respectively act as sources (+) and drains (−) of the Berry curvature (green arrows)[15].

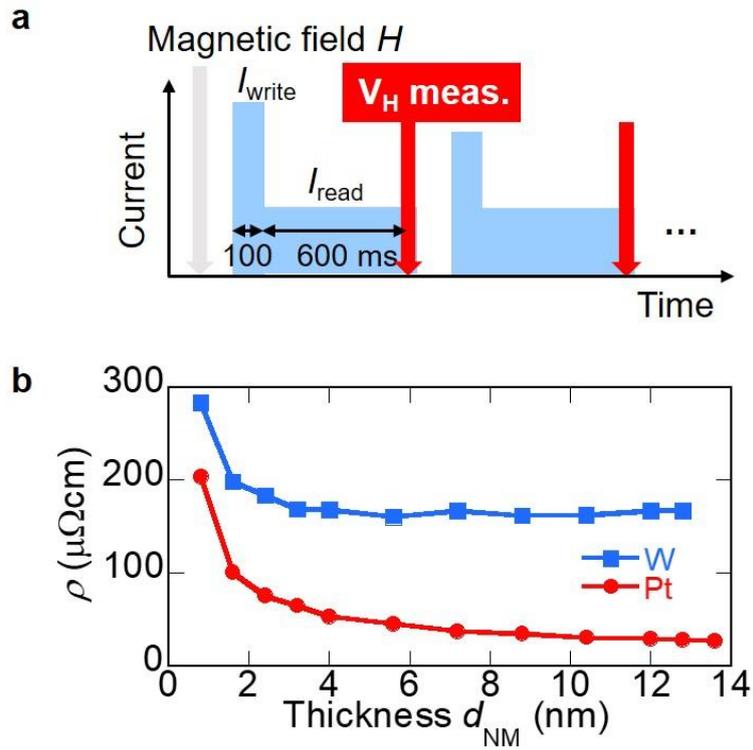

**Extended Data Figure 6 | Experimental conditions for electrical measurements. a,** Schematic diagram of the sequence used for the SOT induced switching measurements. **b,** Thickness dependence of the resistivity of the NM (Pt or W) layer obtained in the Si/SiO$_2$/Ru(2)/Mn$_3$Sn(40)/Pt or W ($d_{NM}$)/AlO$_x$(5) Hall bar devices at room temperature.

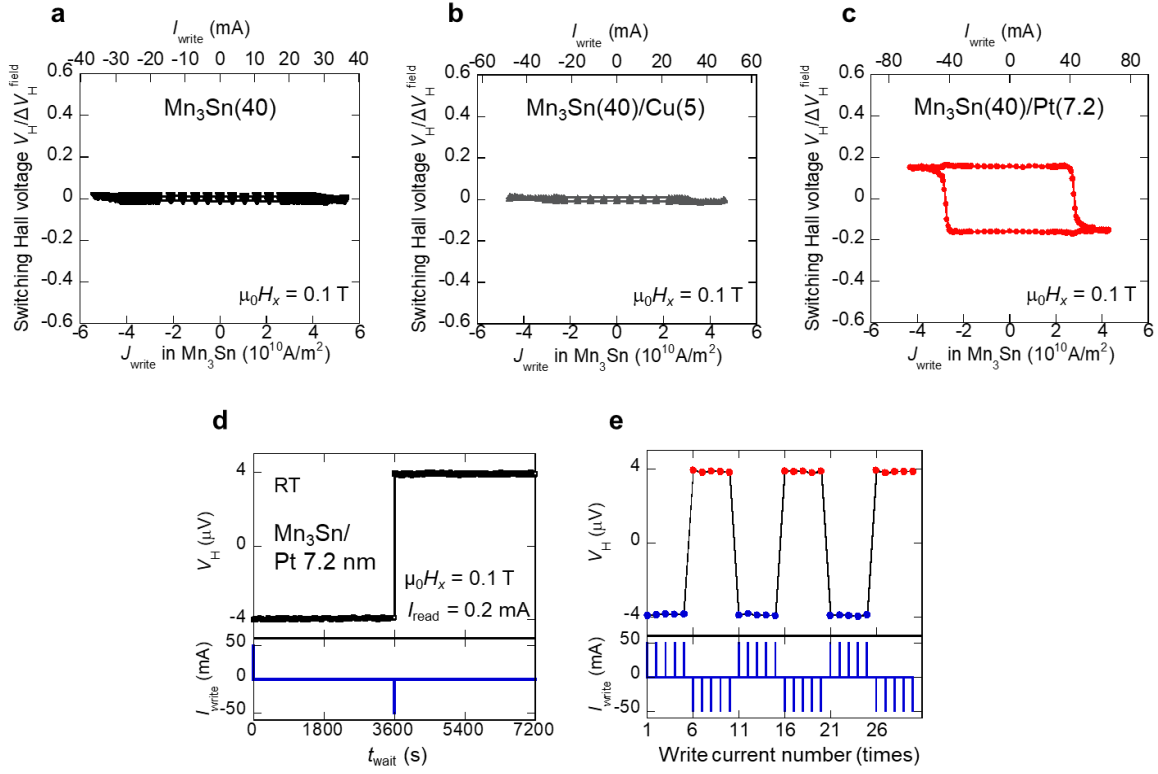

**Extended Data Figure 7 | Current-induced switching, signal stability, and heating effects in the Mn3Sn devices. a-c,** Hall voltage vs. write current (density) for the Mn3Sn without a non-magnetic metal layer (**a**), Mn3Sn/Cu (**b**), and Mn3Sn/Pt (**c**) Hall bar devices under the bias field of $H_x = 0.1$ T. In contrast to the Mn3Sn/Pt device (**c**) showing the clear switching, the Hall voltage of the Mn3Sn sample (**a**) is not switched by the electric current, similarly to the observation made for the Mn3Sn/Cu sample (**b**). Data used in Extended Data Fig. 7**b** and 7**c** are adapted from Fig. 1**g** in the main text. The $x$-axis at the top and bottom of each panel presents the write current $I_{write}$ in whole multilayers and write current density $J_{write}$ in Mn3Sn layer, respectively. **d,** Wait time $t_{wait}$ dependence of the Hall voltage $V_H$ measured after the electrical switching of AHE by the write current $I_{write}$ (±50 mA, 100 ms) in the Mn3Sn/Pt(7.2 nm) device at room temperature. No variation of the AHE signal for $t_{wait}$ = 600 ms ~ 1 hour is observed, which indicates that 600 ms is long enough to cool the sample down to room temperature and the AHE signal is very stable in the Mn3Sn Hall bar devices after the electrical switching. **e,** Write current number dependence of the Hall voltage in Mn3Sn/Pt(7.2 nm) device at room temperature. The AHE signal obtained after the first write current (±50 mA, 100 ms) does not change even after the total five consecutive pulses, similar to the case in the FM systems[31].

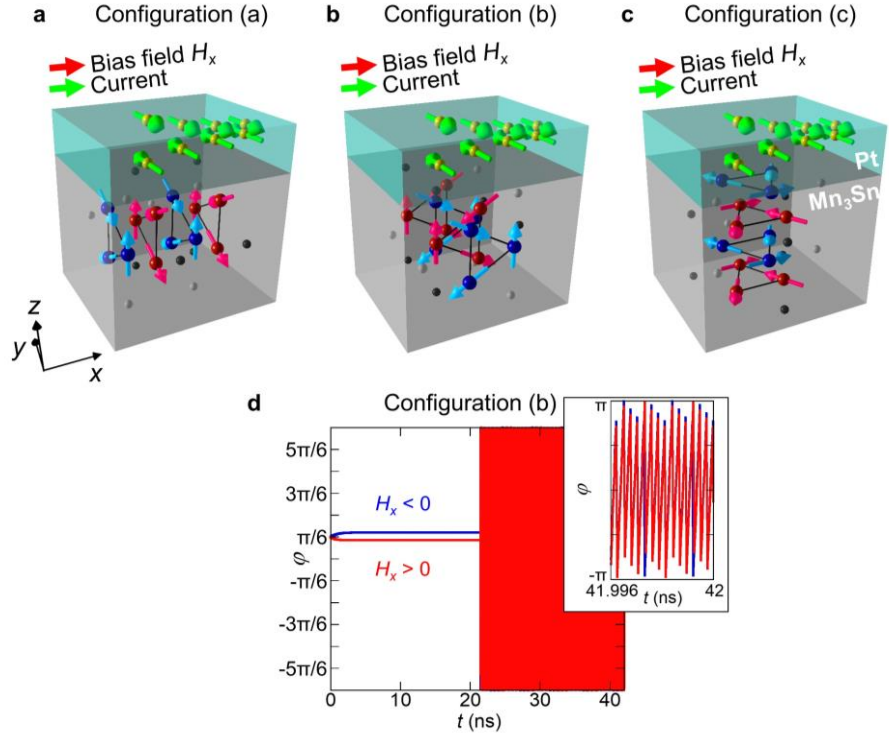

**Extended Data Figure 8 | Crystal grain configurations in the polycrystalline Mn$_3$Sn layer. a-c,** Schematic configurations of the SOT induced switching in the Si/SiO$_2$/Ru(2)/Mn$_3$Sn(40)/Pt or W ($d_{NM}$)/AlO$_x$(5) Hall bar devices in the configuration (a): **a,** the kagome-layer ⊥ the current $I$ and ∥ the electrically injected carrier spin polarization $p$ (the same figure as Fig. 4**a** in the main text), the configuration (b): **b,** the kagome-layer ∥ $I$ and ⊥ $p$, and the configuration (c): **c,** the kagome-layer ∥ $I$ and ∥ $p$. Green cylindrical arrows represent the spin polarized current in NM (e.g. Pt) induced by the writing current along the $x$-direction. Crystal and magnetic structures of Mn$_3$Sn are presented, following the schematic pictures shown in Figs. 1**a** and 1**b** in the main text. **d,** In-plane angle $\varphi$ (as defined in Fig. 4**c**) of the octupole moment as a function of time $t$ in the case for the configuration (b) namely, $p \parallel y$ and $H \parallel x$ (Extended Data Fig. 8**b**). Here, the coordinates $x$, $y$, and $z$ are defined as in Extended Data Fig. 8**a**. The damping-like torque is applied at $t > 21.5$ ns. Inset: Magnified view of the in-plane angle $\varphi$ of the octupole moment as a function of time. The results indicate continuous rotation of a slightly canted ITS structure with the frequency of ~ 3.5 THz.

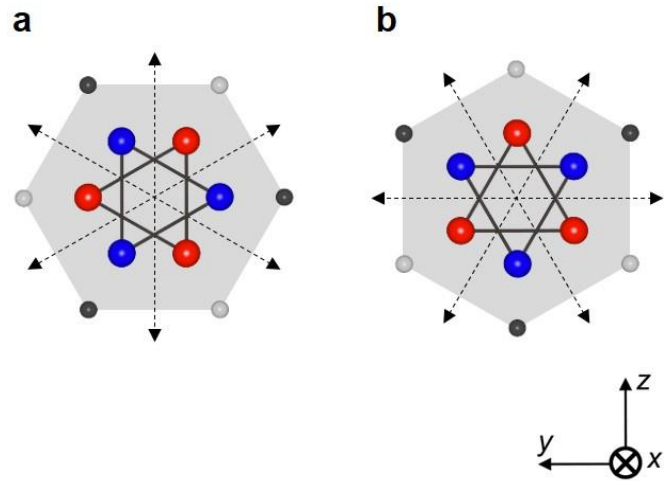

**Extended Data Figure 9 | Kagome-plane arrangements among the configuration (a). a,b,** Schematic figures for the kagome-plane orientations (a-1): the *b*-axis:[$01\bar{1}0$] || *y* (**a**) and (a-2): the *a*-axis:[$2\bar{1}\bar{1}0$] || *y* (**b**) among the configuration (a) in Extended Data Figure 8, where the kagome-layer ⊥ the current *I* and || *y*. The broken arrows represent the magnetic easy-axis for the octupole polarization.

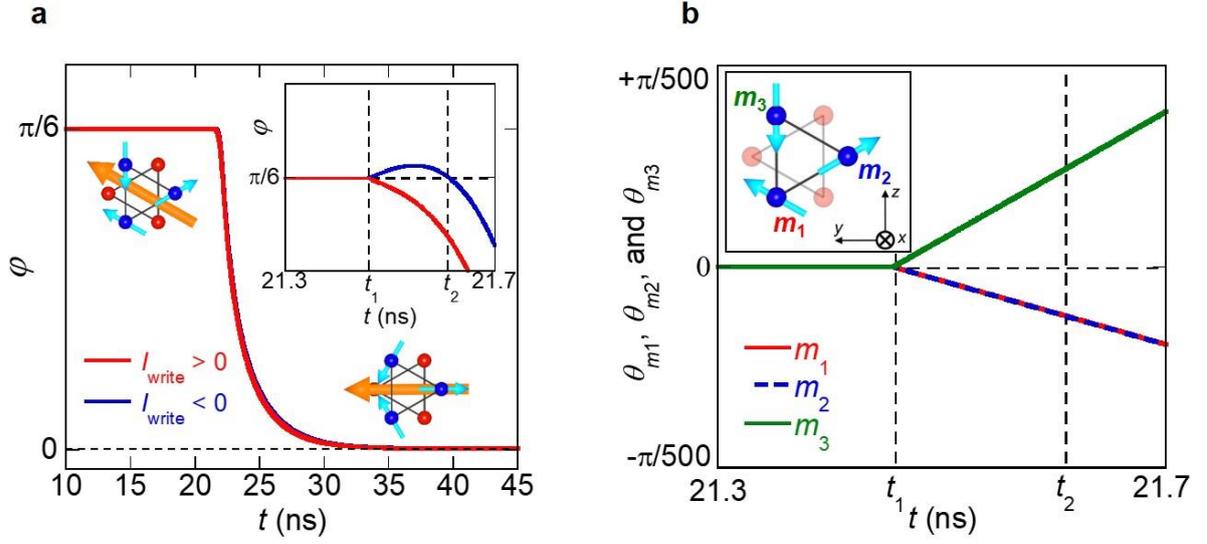

**Extended Data Figure 10 | Simulated dynamics for the sublattice-moments during the switching**

**a,** In-plane motions of the octupole polarization in the absence of the bias field. The red (blue) line corresponds to the motion under $I_{\text{write}} > 0$ ($I_{\text{write}} < 0$). Here, we use the write current with the finite rise time to suppress the incoherent oscillating behavior. The inset shows the enlarged view for the short period right after the current injection at $t_1$. The parameters used here are the same as those used in Fig. 4b in the main text. **b,** Evolution of the out-of-(kagome-)plane components (∥ the $x$-direction) of the sublattice magnetic moments, $m_1$ (red), $m_2$ (blue), and $m_3$ (green), induced by $I_{\text{write}} < 0$. $\theta_{ma}$ ($a = 1, 2,$ and $3$) $> 0$ ($\theta_{ma} < 0$) corresponds to the positive (negative) component to the $x$-direction from the $yz$-(kagome-)plane.